\documentclass{PoS}
\usepackage{graphicx}
\usepackage{amssymb}
\usepackage{amsmath}
\usepackage{mathrsfs}
\usepackage[utf8]{inputenc}
\usepackage{xcolor}
\usepackage{cite}
\definecolor{pb}{RGB}{0, 100, 230}

\title{The Light-Quark Magnetic Moment of the Lambda(1405) Antikaon-Nucleon 
Molecule}

\ShortTitle{The Light-Quark Magnetic Moment of the Lambda(1405) 
Antikaon-Nucleon Molecule}

\newcommand{\CSSM}{Special Research Centre for the Subatomic Structure
  of Matter (CSSM),\\Department of Physics, University of
  Adelaide, Adelaide, South Australia 5005, Australia} 

\newcommand{\CoEPP}{ARC Centre of Excellence for Particle Physics at
  the Terascale (CoEPP),\\Department of Physics, University
  of Adelaide, Adelaide, South Australia 5005, Australia} 

\newcommand{\NCI}{National Computational Infrastructure
  (NCI),\\Australian National University, Canberra, Australian Capital Territory
  0200, Australia}

\author{\speaker{Jonathan M. M. Hall}${}^{a}$, Waseem Kamleh${}^a$, Derek B. 
Leinweber${}^a$, Benjamin J. Menadue${}^{a,b}$, Benjamin J. Owen${}^{a}$, 
Anthony W. Thomas${}^{a,c}$\\
${}^a$\CSSM
\\
${}^b$\NCI\\
${}^c$\CoEPP\\
E-mail: \email{jonathan.hall@adelaide.edu.au}}

\abstract{
The light-quark sector of the $\Lambda(1405)$ baryon is examined in the context 
of the recent discovery of a dominant antikaon-nucleon composition at low quark 
masses. Further evidence for this interpretation of the $\Lambda(1405)$ is 
presented, by calculating the $u$ and $d$ quark contributions to the 
$\Lambda(1405)$ magnetic form factors in lattice QCD. The extent to which these 
quantities are consistent with the exotic molecular description can then be 
quantified by comparing the results with the equivalent nucleon form factors. 
Drawing on a recent extension of the graded-symmetry approach for the 
flavor-singlet components of the $\Lambda(1405)$, the separation of the 
connected and disconnected contributions is performed in both the flavor-octet 
and singlet representations. In both cases, the disconnected loop contributions 
are found to be unexpectedly large.
The relationship between the light-quark contributions to the
$\Lambda(1405)$ magnetic form factor and the connected contributions
of the nucleon magnetic form factors is thus confirmed in the case of lattice 
QCD, establishing compelling evidence for a $\overline{K}N$ molecular structure 
of the $\Lambda(1405)$ near the physical point.}

\FullConference{The 26th International Nuclear Physics Conference\\
		 11-16 September, 2016\\
		 Adelaide, Australia}

\begin{document}

\section{Introduction}
\label{sect:intro}

Efforts to understand the internal structure of the $\Lambda(1405)$ baryon has 
presented a challenge to the scientific community for over 50 years 
\cite{Dalitz:1960du,Dalitz:1967fp}. While a simple quark-model picture 
suggests a negative-parity excited state comprising \textit{up}, \textit{down} 
and \textit{strange} quarks should be heavier 
than the corresponding negative-parity state of the nucleon, its unexpectedly 
light mass has prompted a wide range of studies that strongly indicate the 
presence of a significant contribution from a $\overline{K} N$ bound state 
\cite{Veit:1984an, Kaiser:1995eg,Oset1998, Geng:2007hz, Guo2013, 
Molina:2015uqp,Oller:2013zda, Hall:2014uca, Hall:2014gqa}.
This molecular picture of the structure of the $\Lambda(1405)$ has been shown 
to be prevalent at light quark masses \cite{ Oller:2013zda,Molina:2015uqp}, 
while lattice QCD continues to find compelling evidence for the 
constituent-quark model of the $\Lambda(1405)$ at both intermediate and heavy 
quark masses \cite{Menadue:2011pd,Hall:2014uca}. 
Describing the behaviour of the $\Lambda(1405)$ over a wide range of quark 
masses has recently been shown to require 
a composite picture, whereby a three-quark state with non-zero bare mass 
encounters strong interactions with nearby multi-hadron states, 
particularly $\overline{K} N$ and $\pi\Sigma$, which contribute significantly 
to the properties of the $\Lambda(1405)$ near the physical point 
\cite{Hall:2013qba,Hall:2014uca,Liu:2016wxq}. This development serves to 
bridge the constituent-quark picture at heavy quark
masses and the molecular $\overline{K} N$ dominance of the $\Lambda(1405)$ at 
light quark masses.

While lattice QCD simulation methods have demonstrated unprecedented accuracy 
in determining baryon ground state observables at relatively low quark masses
\cite{Green:2015wqa,Sufian:2016pex},
advancement in excited-baryon form factors is still at an early stage 
\cite{Owen:2013pfa,Roberts:2013oea,Owen:2014txa,Menadue:2013kfi,Hall:2014uca}. 
Nevertheless, calculations of a vanishing strange magnetic form factor for the 
$\Lambda(1405)$ at small quark masses 
are crucial in uncovering how the $\overline{K} N$
molecular behaviour appears on the lattice \cite{Hall:2014uca}. 
Since the $s$ quark present in the $\overline{K} N$ system is confined within 
the spin-$0$ kaon in a relative $S$ wave with respect to a nucleon,  the angular
momentum must be zero. Hence, the strange quark cannot contribute to the 
magnetic form factor of a $\Lambda(1405)$ composed as a molecular $\overline{K}
N$ bound state.

\begin{figure}[t]
\begin{center}
\includegraphics[width=0.5\columnwidth,angle=0]{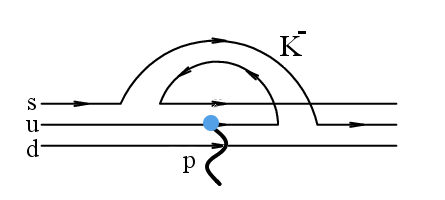}\hspace{-5mm}
\includegraphics[width=0.5\columnwidth,angle=0]{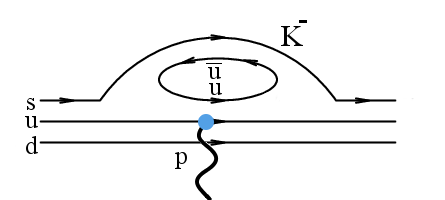}
\includegraphics[width=0.5\columnwidth,angle=0]{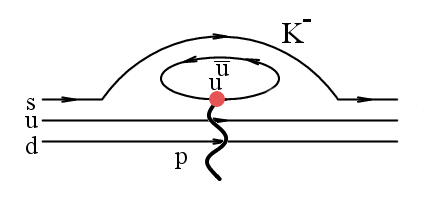}
\vspace*{-8pt}
\caption{(color online). The quark flow diagrams for the process $\Lambda(1405) 
\rightarrow K^-\, p$ can be decomposed into a completely-connected part and two 
parts involving disconnected sea-quark loop contributions.  The upper-left 
completely-connected diagram and the upper-right diagram (\color{pb}
\textit{blue} \color{black} dots) are included in the lattice QCD calculations 
as the photon interacts with a valence quark.  The case where a photon couples 
to a disconnected sea quark loop, illustrated in the lower diagram 
(\color{red}\textit{red} \color{black} dot), is not included in the lattice 
QCD calculations.
\vspace*{-12pt}
\label{fig:dcomp}}
\end{center}
\end{figure}

In the case of the light-quark sector of the $\Lambda(1405)$,   
lattice results must omit photon couplings to
the disconnected quark--antiquark loops, due to computational intensiveness 
and difficulty in isolating a signal for excited states \cite{Mahbub:2013ala,
Morningstar:2013bda,Alexandrou:2014mka,Owen:2015fra,Hall:2013qba,Hall:2014uca,
Leinweber:2015kyz,Liu:2016uzk,Lang:2016hnn}. Diagrams corresponding to the 
included and omitted contributions to the form factors for the process 
$\Lambda^* \to K^-\, p$, for example, are shown in Fig.~\ref{fig:dcomp}. (The
star superscript notation indicates that the resonance has odd parity). 
Thus, careful accounting of the omitted contributions are vital in a 
meaningful comparison of the form factors of the $\Lambda(1405)$ compared to 
those of the nucleon, and a reliable estimate of the magnitude of these partial 
quenching effects must be made \cite{Hall:2016kou}.

\section{Loop contributions to the magnetic form factor}

In the $\overline{K} N$ picture, the spin-$0$ kaon is in a relative $S$ wave 
about the nucleon, and thus, in the absence of orbital angular momentum, the 
light-quark contributions to the magnetic form factor of the
$\Lambda(1405)$ can be attributed solely to the nucleon.  Since couplings 
derived from SU$(3)$ symmetry for 
$\Lambda^* \to K^-\, p$ and $\Lambda^* \to \overline{K}^0 \, n$ are equal, the 
light sector contribution may be written as an average of $n$ and $p$ magnetic 
form factors
\begin{equation}
|\, \Lambda^* \rangle =  \frac{1}{\sqrt{2}} \left ( \, |\, K^- p \rangle + |\, 
\overline{K}^0 n
\rangle \, \right ) \, .
\label{eq:simpleModel}
\end{equation}
In full QCD, which includes the disconnected sea-quark loop contributions, 
the form of the light-quark magnetic form factor takes the form
\begin{eqnarray}
\langle \Lambda^* \,|\, \hat\mu_q \,|\, \Lambda^* \rangle 
&=& \frac{1}{2} \langle K^- p \,|\, \hat\mu_q \,|\, K^- p \rangle + \frac{1}{2} 
\langle \overline{K}^0 n
\,|\, \hat\mu_q \,|\, \overline{K}^0 n \rangle \, , \nonumber \\
&=& \frac{1}{2} \langle p \,|\, \hat\mu_q \,|\, p \rangle + \frac{1}{2} \langle 
n\,|\, \hat\mu_q \,|\,  n \rangle \, .
\label{eq:nucleonRelation}
\end{eqnarray}

While the $\Lambda(1405)$ is identified as the low-lying flavor-singlet
baryon in the $SU(3)$-flavor limit, as one approaches the physical regime, 
significant mixing with octet-flavor symmetry is 
encountered~\cite{Menadue:2011pd}. Therefore one needs to consider both
flavor-octet and flavor-singlet couplings for $\Lambda^* \rightarrow 
\overline{K} N$.  In addressing the latter, we draw upon the recently 
developed graded-symmetry approach for singlet baryons \cite{Hall:2015cua}, 
which augments the standard octet-baryon Lagrangian with the necessary 
additional terms.

First, consider the contributions to the singlet component of the 
$\Lambda(1405)$, denoted $\Lambda^{\prime *}$, where the prime indicates that a 
singlet representation is taken.  In the case of the process 
$\Lambda^{\prime*}\rightarrow K^-\, p$, the relevant ghost term in the 
Lagrangian takes the form \cite{Hall:2015cua}
\begin{equation}
- g_s\, \sqrt{\frac{2}{3}}\; \overline{\Lambda}^{\prime *}\, \tilde{K}^-\,
\tilde{\Lambda}^+_{p,\tilde{u}} \, ,
\end{equation}
where $g_s$ is taken to be the coupling of the singlet to octet-octet process 
$\Lambda^{\prime *}\rightarrow \pi_0\, \Sigma_0$.  Here, the notation of 
Ref.~\cite{Hall:2015cua} is used. $\tilde{K}^-$ is composed of a strange quark 
($s$) and a ghost anti-up quark ($\overline{\tilde{u}}$) and 
$\tilde{\Lambda}^+_{p,\tilde{u}}$ represents a proton-like particle
composed of $\tilde{u}\,u\,d$, with the normal quarks in an anti-symmetric 
formation.  The factor $\sqrt{2/3}$ is derived from the $SU(3|3)$ symmetry 
relations that govern the augmented Lagrangian. Recalling the full QCD vertex 
required for this process,
\begin{equation}
g_s\, \overline{\Lambda}^{\prime *}\, K^-\, p \, ,
\end{equation}
the relative sizes of the connected and disconnected loop contributions can be
resolved-  the connected diagram
has weight $(1/3)\, g_s^2$ and the disconnected diagram has weight
$(2/3)\, g_s^2$. The same ratio is found in the case of $\Lambda^{\prime *}
\rightarrow \overline{K}^0\, n$, where a $d$ quark participates in the
loop in full QCD. 

Since flavor-symmetry breaking leads to significant singlet and octet mixing in 
the flavor components of the $\Lambda(1405)$, 
particularly near the physical point \cite{Menadue:2011pd}, one must also 
consider the octet-to-octet meson and baryon contributions.  Upon partial 
quenching, the corresponding couplings lead to the same ratio of $\sqrt{2/3}$. 
Thus the connected diagram holds a weight of $1/3$ and the disconnected diagram 
holds a weight of $2/3$ of the full QCD process  
independent of the representation of the $\Lambda(1405)$. 
As a result, the calculation of the partially quenched value of the magnetic 
form factor is straightforward.
For example, because half of the $u$-quark contributions are in the 
disconnected loop, one can obtain the $u$-quark contributions to the proton, 
$u_p$, by subtracting off $1/2\times 2/3 = 1/3$ of the full QCD contribution. 
The $u$-quark contribution to the neutron, $u_n$, on the other hand, is fully 
included in the lattice QCD calculation, since the disconnected
quark-loop flavor is a $d$ quark, not a $u$ quark, so no adjustment is required
\begin{align}
&\langle \Lambda^* \,|\, \hat\mu^{\rm conn}_u \,|\, \Lambda^* \rangle 
= \nonumber\\
&\frac{1}{2}\, \left (
\langle K^- p \,|\, \hat\mu_u \,|\, K^- p \rangle 
- \frac{1}{2}\, \frac{2}{3} \, \langle K^- p \,|\, \hat\mu_u \,|\, K^- p \rangle
\right ) 
+ \frac{1}{2}\, \langle \overline{K}^0 n \,|\, \hat\mu_u \,|\, \overline{K}^0 n 
\rangle \, 
= \frac{1}{2}\, \left ( 2 u_p - \frac{2}{3} u_p  + u_n \right )\, .
\label{eqn:muu}
\end{align}
The first two terms of Eq.~(\ref{eqn:muu}) represent the connected
$u$-quark contribution from the proton component within the $\Lambda(1405)$.  
The first of these terms provides the full QCD contribution while the second 
term subtracts half of the weight of the disconnected sea-quark loop associated 
with photon couplings to the disconnected loop. 
Similarly, for the $d$-quark contribution, a magnetic form factor of 
$\frac{1}{2}( 2 d_n - \frac{2}{3} d_n  + d_p)$ is obtained. 
Note that, under charge symmetry, the two light quark contributions, 
$\hat\mu^{\rm conn}_\ell$ become equal for quarks of unit charge
\begin{equation}
\langle \Lambda^* \,|\, \hat\mu^{\rm conn}_\ell \,|\, \Lambda^* \rangle 
= \frac{1}{2}\, \left ( 2 u_p - \frac{2}{3} u_p  + u_n \right )\, .
\label{eqn:finalModel}
\end{equation}

\section{Lattice QCD results}

To test the $\overline{K} N$ model prediction of Eq.~(\ref{eqn:finalModel}), 
the same set of configurations explored in Ref.~\cite{Hall:2014uca} are used, 
leading to a calculation of the left-hand side of the equation,
$\langle \Lambda^* \,|\, \hat\mu^{\rm conn}_\ell \,|\, \Lambda^* \rangle$.
The calculation is based on the $32^3 \times 64$ full-QCD
ensembles created by the PACS-CS collaboration~\cite{Aoki:2008sm},
made available through the International Lattice Data Grid (ILDG)
\cite{Beckett:2009cb}. The ensembles provide a lattice volume of
$(2.9\ \mbox{fm})^3$ with five different masses for the light $u$ and
$d$ quarks, and constant strange-quark simulation parameters.  A
 valence strange quark with a hopping parameter of
$\kappa_s = 0.13665$ is simulated, reproducing the correct kaon mass in the
physical limit \cite{Menadue:2012kc}. The squared pion mass is chosen as
a renormalization group invariant measure of the quark mass, and the
scale is set via the Sommer parameter \cite{Sommer:1993ce} with $r_0 =
0.492$ fm \cite{Aoki:2008sm}.

\begin{figure}[t]
\centering
\includegraphics[height=0.7\hsize,angle=90]{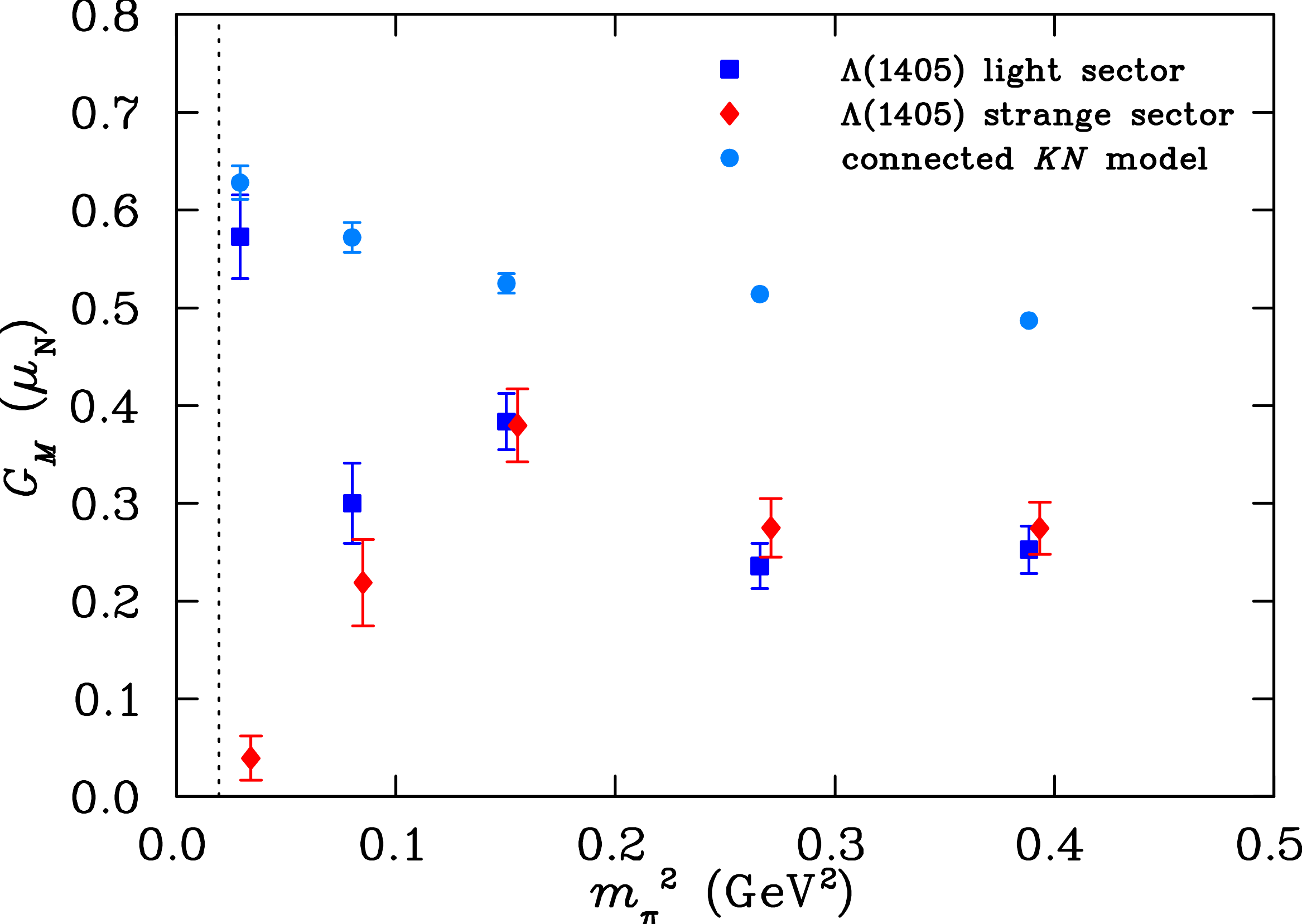}
\vspace{-1mm}
\caption{\footnotesize{The light ($u$ or $d$) and strange quark contributions 
to the magnetic form factor of the $\Lambda(1405)$ at $Q^2 \simeq 0.16$ 
GeV${}^2/c^2$ from Ref.~\cite{Hall:2014uca} are presented as a function of the 
light $u$- and $d$-quark masses, indicated by the squared pion mass, $m_\pi^2$.  
Sector contributions are for single quarks of unit charge.  
The lattice calculations are compared to the predictions of the connected 
$\overline{K} N$ model developed herein. The vertical dashed line indicates the 
physical pion mass. The strange form factor results are offset a small amount 
from the light sector in the $m_\pi^2$-axis for clarity.}
}
\label{fig:mpisq}
\end{figure}

In calculating the right-hand side of Eq.~(\ref{eqn:finalModel}), 
the nucleon magnetic form factors are determined on these lattices using the 
methods introduced in Ref.~\cite{Leinweber:1990dv,Boinepalli:2006xd}, providing 
values of $u_p = 1.216(17)\;\mu_N$ and $u_n = -0.366(19)\;\mu_N$ 
at the lightest pion mass, and the lowest nontrivial momentum transfer of 
$Q^2 \simeq 0.16$ GeV${}^2/c^2$.
 Figure~\ref{fig:mpisq} shows the results for the light- and
strange-quark magnetic form factors of the $\Lambda(1405)$, plotted
as a function of pion mass.  The flavor symmetry present at heavy quark masses 
is broken as the $u$ and $d$ masses approach the physical point, where the
strange magnetic form factor drops nearly to zero.  The light quark
sector contribution differs significantly from the molecular
$\overline{K} N$ model prediction until the lightest quark mass is
reached.  The direct matrix element calculation, $\langle
\Lambda^* \,|\, \hat\mu^{\rm conn}_\ell \,|\, \Lambda^* \rangle$ 
 agrees with the prediction of the connected
$\overline{K} N$ model, shown in Eq.~(\ref{eqn:finalModel}).  At the lightest 
pion mass, the light-quark magnetic form factor of the $\Lambda(1405)$ is 
\cite{Hall:2014uca}
\begin{equation}
\langle \Lambda^* \,|\, \hat\mu^{\rm conn}_\ell(Q^2) \,|\, 
\Lambda^* \rangle  = 0.58(5)\ \mu_N \, ,
\end{equation}
at $Q^2 \simeq 0.16$ GeV${}^2/c^2$.  The connected $\overline{K} N$ model of
Eq.~(\ref{eqn:finalModel}) predicts
\begin{equation}
\langle \Lambda^* \,|\, \hat\mu^{\rm conn}_\ell(Q^2) \,|\, \Lambda^* \rangle 
= 0.63(2)\ \mu_N \, .
\label{eqn:finalPrediction}
\end{equation}
This agreement confirms that the
$\Lambda(1405)$ observed in lattice QCD  is dominated by a molecular 
$\overline{K} N$ structure near the physical point.  

Note that there is a significant shift in the prediction due to the omission of 
the photon couplings to the disconnected sea-quark loop.  In the case where 
such couplings are included, the prediction of the $\overline{K} N$ model is 
much larger, $\langle \Lambda^* \,|\,\hat\mu_\ell \,|\, \Lambda^* 
\rangle = (2\, u_p + u_n) / 2 = 1.03(2)\ \mu_N$. Thus, a determination of these 
disconnected-loop contributions directly from lattice QCD would be of great 
value as a next step forward, particularly for resonances where coupled-channel 
dynamics play an important role.

\newpage
\section{Conclusion}
\label{sect:conc}

The light-quark sector contributions to the magnetic form factor of
the $\Lambda(1405)$, calculated in lattice QCD,  have
been compared to the predictions of a molecular $\overline{K} N$ model. 
By identifying the quark-flow connected contributions to the magnetic form
factors of the $\Lambda(1405)$,  a quantitative analysis of
light-quark contributions to the form factors shows that they are consistent 
with a molecular bound-state description near the physical point.

Identification and removal of the quark-flow disconnected
contributions to the $\overline{K} N$ model were conducted 
using a recently extended graded-symmetry approach~\cite{Hall:2015cua}, 
which includes couplings to the flavor-singlet components of baryons. 
It is found that the ratio of connected
to disconnected contributions is identical for both
flavor-singlet and flavor-octet representations of the $\Lambda(1405)$.  

Using new results for the magnetic form factors of the nucleon at a
light pion mass of $m_\pi = 156$ MeV \cite{Hall:2016kou}, the connected
$\overline{K} N$ model predicts a light-quark sector contribution to
the $\Lambda(1405)$ of $0.63(2)\ \mu_N$, which is consistent 
with the direct calculation of $0.58(5)\ \mu_N$ from
Ref.~\cite{Hall:2014uca}. This investigation thus provides further evidence 
that the internal structure of the $\Lambda(1405)$ is dominated by a 
$\overline{K} N$ molecule.

\begin{acknowledgments}

We thank the PACS-CS Collaboration for making their $2+1$ flavor configurations 
available and the ongoing support of the ILDG.  This research was undertaken 
with the assistance of the University of Adelaide's Phoenix cluster and 
resources at the NCI National Facility in Canberra, Australia.  NCI
resources were provided through the National Computational Merit Allocation 
Scheme, supported by the Australian Government and the University of Adelaide 
Partner Share.  This research is supported by the Australian Research Council 
through the ARC Centre of Excellence for Particle Physics at the
Terascale (CE110001104), and through Grants No.\ LE160100051, DP151103101 
(A.W.T.), DP150103164, DP120104627 and LE120100181 (D.B.L.).

\end{acknowledgments}

\bibliographystyle{h-physrev4}
\bibliography{refs}

\end{document}